\documentclass[12pt]{article}
\title{\bf Metric Unification of Gravitation and Electromagnetism Solves the
Cosmological Constant Problem}
\author{  Murat \"Ozer \\
Department of Physics, College of Science,
King Saud University,\\
P. O. Box 2455, Riyadh 11451, Saudi Arabia}
\newcommand{\be}{\begin{equation}}
\newcommand{\ee}{\end{equation}}
\begin{document}
\maketitle

\begin{abstract}
\noindent We first review the cosmological constant problem, and then
mention a conjecture of Feynman according to which the general relativistic
theory of gravity should be reformulated in such a way that energy does
not couple to gravity. We point out that our recent unification of 
gravitation and electromagnetism through a symmetric metric tensor has
the property that free electromagnetic energy and the vacuum energy
do not contribute explicitly to the curvature of spacetime 
just like the free gravitational energy. Therefore in this formulation of 
general relativity, the vacuum energy density has its very large value 
today as in the early universe, while the cosmological constant does not 
exist at all.
\end{abstract}\vspace{0.5cm}
The mysterious cosmological constant\footnote{See references \cite{abb} for
 a nontechnical and \cite{wei1, car, sah}  for technical reviews.} has 
been with us for
 more than eighty years ever since Einstein introduced it in 1917 \cite{ein}
to obtain a static universe by modifying his field equations to
\be 
R_{\mu\nu}-\frac{1}{2}Rg_{\mu\nu}+\lambda_bg_{\mu\nu}=\frac{8\pi G}
{c^4}T_{\mu\nu},
\ee
where $\lambda_b$ is the (bare) cosmological constant, $R_{\mu\nu}$
is the Ricci tensor, $R=R^{\mu}_{\mu}$ is the curvature scalar, $G$ is 
Newton's
gravitational constant and $c$ is the speed of light. Here $T_{\mu\nu}$ is
the energy-momentum tensor which represents the energy content of space to 
which all forms of energy such as matter energy, radiation energy, 
electromagnetic energy, thermal energy, etc. contribute. Gravitational 
energy, however, is not included in $T_{\mu\nu}$ because it is already
included (implicitly) on the left of eq.(1). For a homogeneous and isotropic
universe $T_{\mu\nu}$ necessarily takes the form of the energy-momentum
tensor of a perfect fluid
\be 
T_{\mu\nu}=\left(\rho+\frac{p}{c^2}\right)U_{\mu}U_{\nu}+
\frac{p}{c^2}g_{\mu\nu},
\ee
where $\rho$, $p$, and $U_\mu$ are respectively the energy density, pressure
and four-velocity of the fluid. When the field equations (1) are applied
to the Robertson-Walker metrics
\be 
ds^2=-c^2dt^2+a(t)^2\left[\frac{dr^2}{1-kr^2}+r^2(d\theta^2+sin^2\theta 
d\phi^2)\right]
\ee
the Friedmann equation ensues:
\be 
\left(\frac{\dot a}{a}\right)^2=\frac{8\pi G}{c^4}\rho+\frac{\lambda_bc^2}
{3}-\frac{kc^2}{a^2},
\ee
where $a(t)$ is the scale factor of the universe and $k=-1, 0, +1$
for a universe that is respectively spatially open, flat, and closed.

Be it defined as `empty space' or `the state of lowest energy' of a
theory of particles, the vacuum has a lot of energy associated with it.
The energy of the vacuum per unit volume, the vacuum energy density, has 
numerous sources. Virtual fluctuations of each quantum field corresponding
to a particle and the potential energy of each field contribute to it.
Stipulating that the vacuum be Lorentz invariant entails the energy-momentum
tensor for it to be
\be 
T^{vac}_{\mu\nu}=-\rho_{vac}g_{\mu\nu},
\ee
because $g_{\mu\nu}$ is the only $4\times 4$ tensor that is invariant
under Lorentz transformations. Comparing eq.(5) with eq.(2) reveals that
the vacuum has the equation of state $p_{vac}/c^2=-\rho_{vac}$ 
\cite{zel}. According
to our current understanding, the value of the density of energy that 
resides
in the vacuum has no relevance in nongravitational physics both at the
classical and quantum levels. However, being a form of energy density,
$\rho_{vac}$ takes its place in the field equations (1), thus modifying
$T_{\mu\nu}$ to
\be 
T_{\mu\nu}=T^M_{\mu\nu}+T^{vac}_{\mu\nu},
\ee
where now $T^M_{\mu\nu}$ is the total energy-momentum tensor of the space 
other than that of the vacuum. With $T^{vac}_{\mu\nu}$ included, the 
Friedmann equation (4) changes to
\be 
\left(\frac{\dot a}{a}\right)^2=\frac{8\pi G}{c^4}\rho+
\frac{8\pi G}{c^4}\rho_{vac}+\frac{\lambda_bc^2}
{3}-\frac{kc^2}{a^2},
\ee
from which the `effective cosmological constant' is defined to be
\begin{eqnarray}
\lambda^{eff} &=&\lambda_b+\lambda_{vac}\nonumber\\
 &=&\lambda_b+\frac{8\pi G}{c^4}\rho_{vac} \nonumber\\
 &=&\frac{8\pi G}{c^4}\left(\frac{c^4\lambda_b}{8\pi G}+\rho_{vac}\right)
\nonumber\\
 &=&\frac{8\pi G}{c^4}\rho^{eff}_{vac}.
\end{eqnarray}
This means that even if the bare cosmological constant $\lambda_b$ is zero,
the effective cosmological constant is not. Anything that contributes
to the energy density of the vacuum is tantamount to a cosmological constant.
The present value of $\rho^{eff}_{vac}$ can be estimated from astronomical
observations. The Hubble constant at the present epoch is $H_0=(\dot a/a)_0=
50-80\:kms^{-1}Mpc^{-1}$. The critical energy density of the universe,
$\rho_c$, in the absence ov $\rho^{eff}_{vac}$ is defined by
\be 
\rho_c=\frac{3H^2_0c^2}{8\pi G}
\ee
and has the value $4.7\times 10^{-27}\:kgm^{-3}\,c^2=2\times10^{-47}\:GeV^4$
for $H_0=50\:kms^{-1}Mpc^{-1}$. In terms of the present values of the
density parameters $\Omega_0=\rho_0/\rho_c$ and $\Omega^{eff}_{vac}=
\rho^{eff}_{vac}/\rho_c$ eq.(7) can be cast into
\be 
1=\left(\Omega_0+\Omega^{eff}_{vac}-\frac{kc^2}{a^2_0H^2_0}\right).
\ee
Since no effects of the spatial curvature are seen, the curvature term in
eq.(10) can be safely neglected. Observations give us that $\Omega_0=0.1
-0.4$, with dark matter included, from which $\Omega^{eff}_{vac}\approx 
0.6-0.9$ follows from eq.(10). Hence, we conclude that today
\begin{eqnarray} 
\rho^{eff}_{vac}<\rho_c\approx 10^{-47}\:GeV^4\nonumber\\
\lambda^{eff}<10^{-52}\:m^{-2}.
\end{eqnarray}

As well known, this is in great disagreement with the values predicted by
gauge field theories, of which the best example and the experimentally
well established one is the electroweak theory \cite{gla, wei2, sal}. In 
this theory, as a result
of spontaneous symmetry breaking with a Higgs doublet $\phi$, the minimum
of the potential of $\phi$ contributes to the vacuum energy density and 
hence to the cosmological constant by \cite{dre,lin}
\begin{eqnarray} 
\mid \rho_{vac}\mid &=&\mid V_{min}\mid =\mid -\frac{1}{8}M^2_Hv^2\mid
\approx2\times 10^8\:GeV^4\nonumber\\
\mid\lambda_{vac}\mid &=&\frac{8\pi G}{c^4}\mid\rho_{vac}\mid\approx
10^3\:m^{-2},
\end{eqnarray}
where it is assumed that the potential vanishes at $\phi=0$, and a Higgs 
particle of mass $M_H\approx 150\:GeV$ together with $v=246\:GeV$ have 
been used. These are larger than the present bounds on $\rho^{eff}_{vac}$
and $\lambda^{eff}$ by a factor of $10^{55}$. The two terms in
$\rho^{eff}_{vac}=c^4\lambda_b/8\pi G+\rho_{vac}$ 
 must cancel to at least 55 decimal places so as to
reduce $\rho^{eff}_{vac}$ and hence $\lambda^{eff}$ to their small values 
today. So, in the context of Einstein's general relativity theory,  the 
cosmological constant problem is to understand through what natural 
mechanism the vacuum energy density $\rho_{vac}$ got reduced to
its small value today. It is not why it was always small. Many different
approaches to the problem, none being entirely satisfactory, have been tried
\cite{wei1}. The idea which triggered the solution that we shall present
here belongs to Feynman. In an interview on Superstrings, while talking about
gravity he said \cite{dav}: `In the quantum field theories, there is an 
energy associated with what we call the vacuum in which everything has 
settled down to the lowest energy; that energy is not zero-according to 
the theory.
Now gravity is supposed to interact with every form of energy and should 
interact then with this vacuum energy. And therefore, so to speak, 
a vacuum would have a weight-an equivalent mass energy-and would produce a 
gravitational field. Well, it doesn't! The gravitational field produced by 
the energy in the electromagnetic field in a vacuum-where there's no light, 
just quiet, nothing-should be enermous, so enermous, it would be obvious. 
The fact is, it's zero! Or so small that it's completely in disagreement 
with what we'd expect from the field theory. This problem is sometimes 
called the cosmological constant problem. It suggests that we're missing 
something in our formulation of the theory of gravity. It's even possible 
that the cause of the trouble-the infinities-arises from the gravity 
interacting with its own energy in a vacuum. And we started off wrong 
because we already know
there's something wrong with the idea that gravity should interact with the
energy of a vacuum. So I think the first thing we should understand is
how to formulate gravity so that it doesn't interact with the energy in a
vacuum. Or maybe we need to formulate the field theories so there isn't any 
energy in a vacuum in the first place.'

The purpose of this letter is to point out that (i) such a formulation of 
gravity
in which the free electromagnetic and vacuum energy does not disturb the
emptiness of space, just like the free gravitational energy, in the sense
that their effects are already implicitly included on the left side of the
field equations has 
recently been formulated by us \cite{ozer1,ozer2} and (ii) there does
not exist a cosmological constant problem in this formulation. This new 
formulation,
however, is not only a formulation of gravity but also a unified description
of gravity and electromagnetism through a symmetric metric tensor. The
impossibility of describing the motion of charged particles with different 
charge-to-mass ratios in an electromagnetic field by a single geometry
leads us to consider classes of geometries corresponding to different
charge-to-mass ratios. This way the electromagnetic force on a given charged
test particle can be geometrized \cite{ozer1}. Consider an object with a  
distribution of charged matter with total mass $M_o$ and charge $Q_o$. Let 
there be another distribution of matter $M$ and charge $Q$ external to the 
object. Let also a test particle of mass $m$ and charge $q$ be moving in 
the
charge distribution external to the object.
The modified field equations are 
\be 
R_{\mu\nu}-\frac{1}{2}Rg_{\mu\nu}=\frac{8\pi G}{c^4}T^M_{\mu\nu}+
\frac{k_e}{c^4}\frac{q}{m}T^{CC}_{\mu\nu},
\ee
as opposed to the Einstein field equations
\be 
R_{\mu\nu}-\frac{1}{2}g_{\mu\nu}R=\frac{8\pi G}{c^4}\left[T_{\mu\nu}^M+
T_{\mu\nu}^{EM}(Q)+ T_{\mu\nu}^{EM}(Q_o)+T^{vac}_{\mu\nu}\right]
\ee
where $k_e$ is the (Coulomb) electric constant, $T^M_{\mu\nu}$
is the matter energy-momentum tensor of
the distribution outside the object. $T_{\mu\nu}^{EM}(Q_o)$ and 
$T_{\mu\nu}^{EM}(Q)$ are the energy-momentum tensors due to the 
electromagnetic fields of the object and the charge distribution outside it,
respectively.
$T^{CC}_{\mu\nu}$ is the so called 
charged-current tensor, given by
\be 
T^{CC}_{\mu\nu}=\frac{1}{3}v_{\alpha}{\cal J}^{\alpha}\left(\frac{1}{c^2}
U_{\mu}U_{\nu}+g_{\mu\nu}\right),
\ee
where $v^{\alpha}=(\gamma_vc,\gamma_v\vec v)$ is the four velocity of the
test particle, ${\cal J}^{\alpha}=(c\rho_{Q}, \vec J+\vec J_D)$ with
$\rho_{Q}$, $\vec J$, and $\vec J_D$ being respectively the charge density,
current density, and the displacement current density of the  charge 
distribution outside the object, $U^{\mu}=(\gamma_uc,\gamma_u
\vec u)$ is the four-velocity of the charge distribution, and 
$\gamma_{v(u)}=(1-v^2(u^2)/c^2)$. $T^{CC}_{\mu\nu}$ is not an energy
momentum tensor. The right-hand side of eq.(13) does not contain the energy-
momentum tensor of the electromagnetic field due to  the charge distribution
of the object. For example, the field equations 
describing a spherical distribution of mass $M$ and charge $Q$ located at
$r=0$ are
\be 
R_{\mu\nu}=0
\ee
and has the solution
\begin{eqnarray} 
ds^2=-\left(1-2\frac{GM}{c^2r}+2\frac{q}{m}\frac{k_eQ}{c^2r}\right)c^2dt^2+
\left(1-2\frac{GM}{c^2r}+2\frac{q}{m}\frac{k_eQ}{c^2r}\right)^{-1}dr^2
\nonumber\\
+r^2d\theta^2+r^2sin^2\theta d\phi^2.
\end{eqnarray}
Whereas in Einstein's general relativity, one has
\be 
R_{\mu\nu}=\frac{8\pi G}{c^4}T^{EM}_{\mu\nu}
\ee
instead of eq.(15), where $T^{EM}_{\mu\nu}$ is the energy-momentum tensor
of the electromagnetic field of the charged sphere. Einstein's general
relativity is a theory of gravitation. Our general relativity is a theory
of gravitation and electromagnetism.
As well known, the solution of eq.(17) is the Reissner-Nordstr{\o}m solution
\begin{eqnarray} 
ds^2=-\left(1-2\frac{GM}{c^2r}+\frac{Gk_eQ^2}{c^4r^2}\right)c^2dt^2+
\left(1-2\frac{GM}{c^2r}+\frac{Gk_eQ^2}{c^4r^2}\right)^{-1}dr^2+\nonumber\\
r^2d\theta^2+r^2sin^2\theta d\phi^2,
\end{eqnarray}

To sum up, this new formulation of general relativity describes gravitation
and electromagnetism  as an effect of the curvature
of spacetime produced by matter energy and  electric charge. 
We have suggested a very simple deflection of electrons by a spherical 
charge distribution experiment in ref.\cite{ozer2} to distinguish
between the two theories.

Having presented the salient features of our formulation, we can now present
our solution to the cosmological constant problem: The vacuum energy today
is as large as it can be. It has the same very large energy density
today as it had in the early universe. The contribution of the vacuum to
the cosmological constant is simply zero. This is because no form of 
energy-momentum tensor except for that of the matter is allowed on the
right side of the modified field equations (13). The effect on the
curvature of spacetime of any form of free energy like gravitational, 
electromagnetic, and vacuum is already included (implicitly)
on the left of eq.(13). In Einstein's general relativity the contribution of
$\rho_{vac}$ to $\lambda^{eff}$ is $\lambda_{vac}=(8\pi G/c^4)\rho_{vac}$, 
whereas in our formulation $\lambda_{vac}=0\times \rho_{vac}=0$. As for the
bare cosmological constant $\lambda_b$ in eq.(1), one may mathematically
include it in our eq.(13). But it cannot have the physical meaning of a
some sort of bare contribution to the effective vacuum energy density. If,
somehow, there is an unknown bare contribution $\rho_b$ to the (effective)
vacuum energy density, then the corresponding bare cosmological constant is
$\lambda_b=0\times \rho_b=0$ in our formulation.

In conclusion, if a metric unification of gravitation and electromagnetism
is realized in nature, then gravitation is no different from other 
interactions so far as the effects of the vacuum energy are concerned. The
vacuum energy density and the cosmological constant are not related; the
former  has a very large value while the latter does not exist at all.
Immediate performance of the experiment of the deflection of electrons by 
a positive spherical charge distribution \cite{ozer2} cannot be 
overemphasized.


\begin{thebibliography}{20}
\bibitem{abb} L. Abbott, Sci. Am. May 1988(1988)82. 
\bibitem{wei1} S. Weinberg, Rev. Mod. Phys. 61(1989)1.
\bibitem{car} S. M. Carroll, W. H. Press, E. L. Turner, Annu. Rev. Astron.
Astrophys. (1992)499.
\bibitem{sah} V. Sahni and A. Starobinsky, astro-ph/9904398.
\bibitem{ein} A. Einstein, Sitz. Preuss. Akad. Wiss. 142(1917).
\bibitem{zel} Ya. B. Zel'dovich, JETP Lett. 6(1967)316; Sov. Phys. Uspekhi
11(1968)381.
\bibitem{gla} S. L. Glashow, Nucl. Phys. 22(1961)579.
\bibitem{wei2} S. Weinberg, Phys. Rev. Lett. 19(1967)1264.
\bibitem{sal} A. Salam, in Proceedings of the VIII Nobel
Symposium, edited by N. Svartholm (Almqvist and Wiksell, Stockholm, 1968), 
p.367.
\bibitem{dre} J. Dreitlein, Phys. Rev. Lett. 33(1974)1243.
\bibitem{lin} A. D. Linde, JETP Lett. 19(1974)183.
\bibitem{dav} P. C. W. Davies and J. Brown (ed.), Superstrings, A Theory
of Everything, (1988) Cambridge University Press, p.201.
\bibitem{ozer1} M. \"Ozer, `On the Equivalence Principle and a Unified
Description of Gravitation and Electromagnetism', gr-qc/9910062.
\bibitem{ozer2} M. \"Ozer, `Proposed Experiments to Test the Unified 
Description of Gravitation and Electromagnetism Through a 
Symmetric Metric', gr-qc/9910095.
\end{thebibliography}
\end{document}